\def\year{2021}\relax
\documentclass[letterpaper]{article} % DO NOT CHANGE THIS
\usepackage{aaai21}  % DO NOT CHANGE THIS
\usepackage{times}  % DO NOT CHANGE THIS
\usepackage{helvet} % DO NOT CHANGE THIS
\usepackage{courier}  % DO NOT CHANGE THIS
\usepackage[hyphens]{url}  % DO NOT CHANGE THIS
\usepackage{graphicx} % DO NOT CHANGE THIS
\usepackage{natbib}  % DO NOT CHANGE THIS AND DO NOT ADD ANY OPTIONS TO IT
\usepackage{caption} % DO NOT CHANGE THIS AND DO NOT ADD ANY OPTIONS TO IT
\usepackage[caption=false]{subfig}
\usepackage{xcolor}
\usepackage{booktabs}
\usepackage{multirow}
\title{Studying Moral-based Differences in the Framing of Political Tweets}  
\author{Markus Reiter-Haas \textsuperscript{\rm 1},
Simone Kopeinik \textsuperscript{\rm 2},
Elisabeth Lex \textsuperscript{\rm 1}\\
}
\affiliations{
    \textsuperscript{\rm 1}Graz University of Technology, \\
    \textsuperscript{\rm 2}Know-Center GmbH, \\
    reiter-haas@tugraz.at, 
    skopeinik@know-center.at, 
    elisabeth.lex@tugraz.at
}

\begin{document}

\maketitle

\begin{abstract}
    
    %%framing bias and intensity
    
   In this paper, we study the moral framing of political content on Twitter. Specifically, we examine differences in moral framing in two datasets: (i) tweets from US-based politicians annotated with political affiliation and (ii) COVID-19 related tweets in German from followers of the leaders of the five major Austrian political parties. Our research is based on recent work that introduces an unsupervised approach to extract framing bias and intensity in news using a dictionary of moral virtues and vices. In this paper, we use a more extensive dictionary and adapt it to German-language tweets. Overall, in both datasets, we observe a moral framing that is congruent with the public perception of the political parties. In the US dataset, democrats have a tendency to frame tweets in terms of care, while loyalty is a characteristic frame for republicans. In the Austrian dataset, we find that the followers of the governing conservative party emphasize care, which is a key message and moral frame in the party's COVID-19 campaign slogan. Our work complements existing studies on moral framing in social media. Also, our empirical findings provide novel insights into moral-based framing on COVID-19 in Austria.

\end{abstract}

\section{Introduction}

%investi- gate moral-based differences in the framing of the content expressed by members and followers of opposing political parties.

Politicians and political campaigns increasingly use social media to connect and communicate with potential voters~\cite{graham2013between}. 
The effectiveness of such communication is influenced by how the message is \emph{framed}~\cite{kusmanoff2020five}. Framing corresponds to the act of changing the formulation of a problem to affect the choices of people \cite{tversky1981framing}. %It is applied in many domains such as politics~\cite{lakoff2014all} or marketing~\cite{grau2007cause}.

%However, identifying framing is a nontrivial task. 
Recently, several related works focus on the characterization of frames: \citet{walter2019news} use topic modeling and network analysis to identify frames in news.  \citet{shurafa2020political} categorize political discussions related to COVID-19 in Twitter into either \emph{blame frames} or \emph{support frames}. \citet{wicke2020framing} find that the discourse around COVID-19 on Twitter is framed using war-related terminology. 

%\citet{kwak2020frameaxis} use an embedding-based approach termed \emph{FrameAxis} to extract the most relevant semantic axes of a text characterized by opposing word pairs. They show that their approach can quantify the extent to which a text is biased toward one of the two opposing words.

% \para{Aim of this work.} 
In our work, we aim to study differences in moral-based framing in content created by members and followers of opposing political parties on Twitter. We base our approach on the work of~\citet{mokhberian2020moral}, who have recently introduced an unsupervised, embedding-based method to characterize \emph{moral frames} in text. Moral frames are frames that emphasize specific moral virtues and vices, such as care or harm. The approach of Mokhberian et al. is grounded in the Moral Foundation Theory from the social sciences~\cite{haidt2004intuitive}, which defines five basic moral foundations and their associated virtues and vices~\cite{haidt2007moral}. Based on the theory, several moral foundation dictionaries~\cite{graham2009liberals,frimer2017moral} have been developed that contain prototypical words for each moral foundation.  

In this paper, we employ a similar approach to Mokhberian et al. However, while they utilize the moral foundation dictionary by~\citet{graham2009liberals}, for our experiments, we use the more recent and more extensive dictionary by~\citet{frimer2017moral}. Besides, we translate the content of that moral foundation dictionary to German using a list of sample translations of positive and negative valence words~\cite{weichselbaum2018implicit} and two sets of word embeddings, i.e., one for English and one for German~\cite{grave2018learning}.

For our study, we create two Twitter datasets. The first dataset contains tweets from US-politicians annotated with political affiliation (democrats vs. republicans). 
The second dataset contains COVID-19-related tweets from the followers of the five major Austrian political parties' leaders in the German language. 
From the tweets, we extract moral frames corresponding to the five moral foundations, their frame bias, i.e., the emphasis towards either virtue or vice, and frame intensity, i.e., the extent to which a frame is used. To study the prevalence of moral frames, we train a logistic regression classifier to predict party affiliation and investigate its coefficients.
% of a classifier that predicts parties. \mrh{predicts parties ist noch nicht verständlich}
% \el{man koennte sich fragen, wie viele frames. also ein frame pro tweet, mehr als  frame pro tweet. spezifizieren.}
% \el{moral frames, as well as their frame bias...} 
% per party by training a classifier \el{hier sollte kurz gesagt werden, was der classifier predicted, im naechsten absatz schreibst du ja, welche moral frames fuer was predictors sind}.

In both datasets, we observe a moral framing congruent with the public perception of the political parties. In the US dataset, high frame intensity on care and fairness are predictors for democrats, while high frame intensity on loyalty and sanctity characterise republicans. In the Austrian dataset, we find a frame bias toward care in the COVID-19-related tweets of the conservative political party leader's followers. We attribute this to the followers' adoption of the conservative COVID-19 slogan's moral framing that stresses caring.

\begin{figure*}[!th]
    \centering
    \parbox{.49\textwidth}{%

        \subfloat[Care Axis]{
        \includegraphics[width=0.48\textwidth]{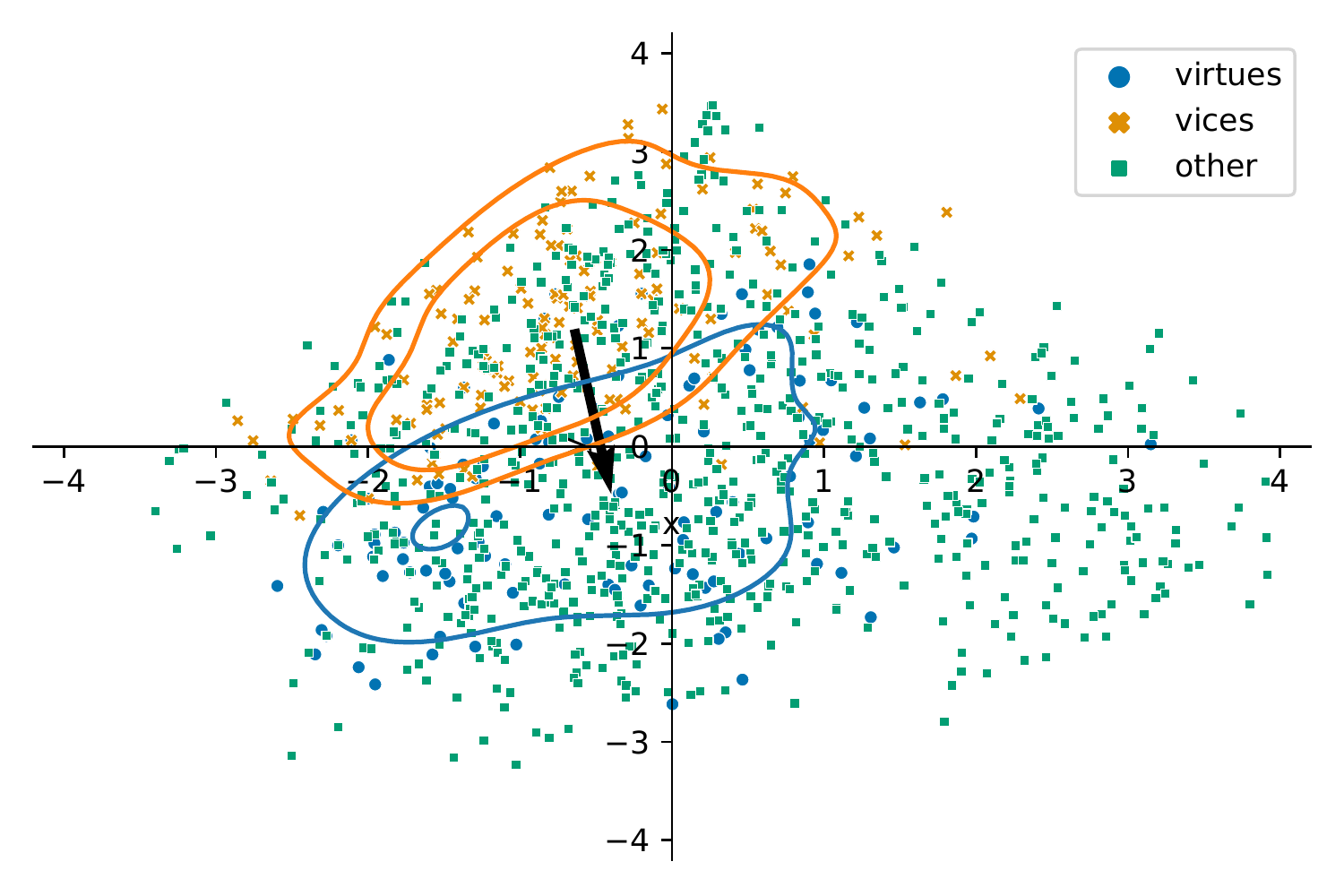}
        }
    }
    \hfill
    \parbox{.5\textwidth}{%
        \begin{tabular}{c@{\hskip -0.05cm}c}
        \subfloat[Fairness Axis]{
        \includegraphics[width=0.23\textwidth]{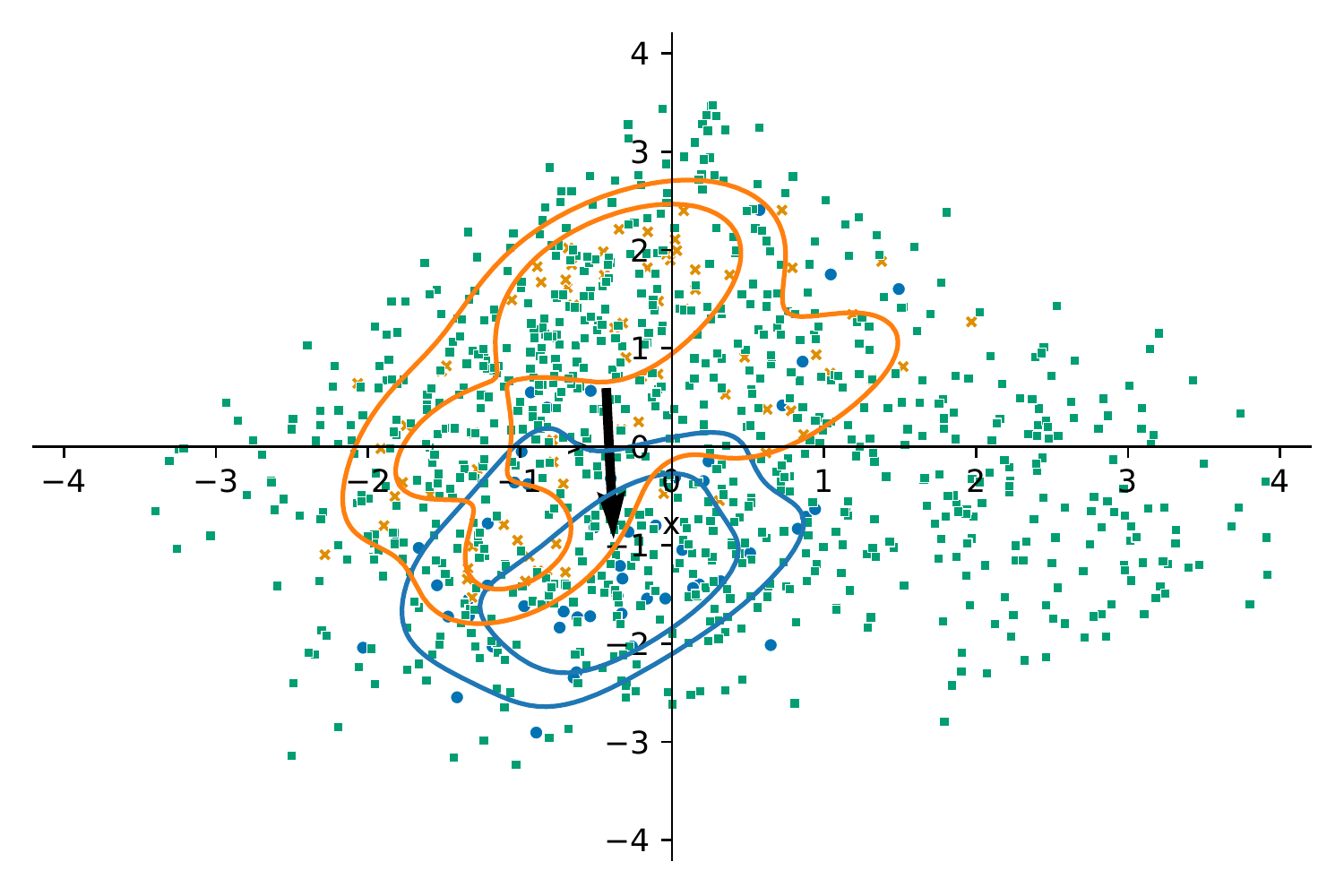}
        }
        &
        \subfloat[Loyalty Axis]{
        \includegraphics[width=0.23\textwidth]{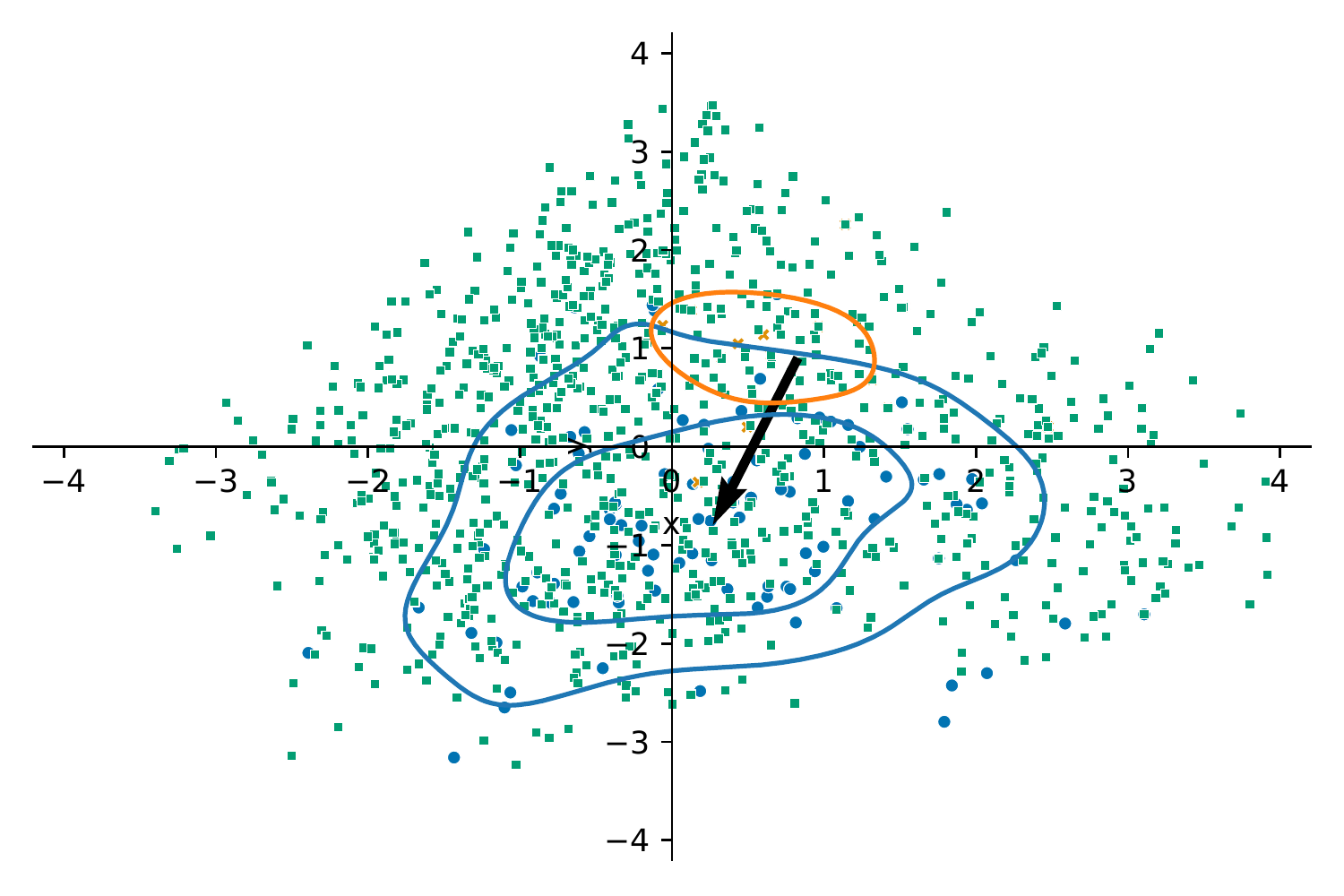}
        } 
        \\
        \subfloat[Authority Axis]{
        \includegraphics[width=0.23\textwidth]{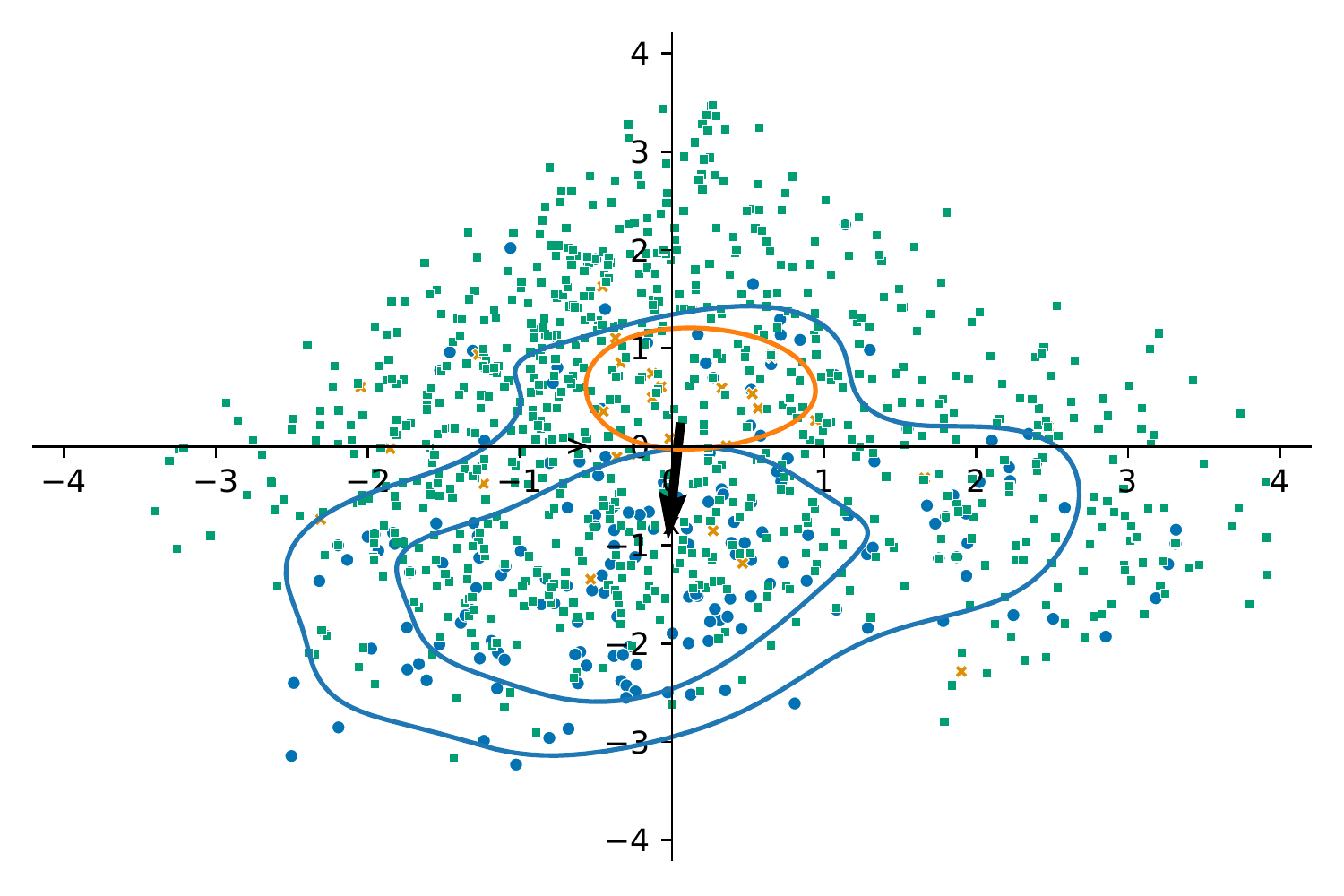}
        }
        &
        \subfloat[Sanctity Axis]{
        \includegraphics[width=0.23\textwidth]{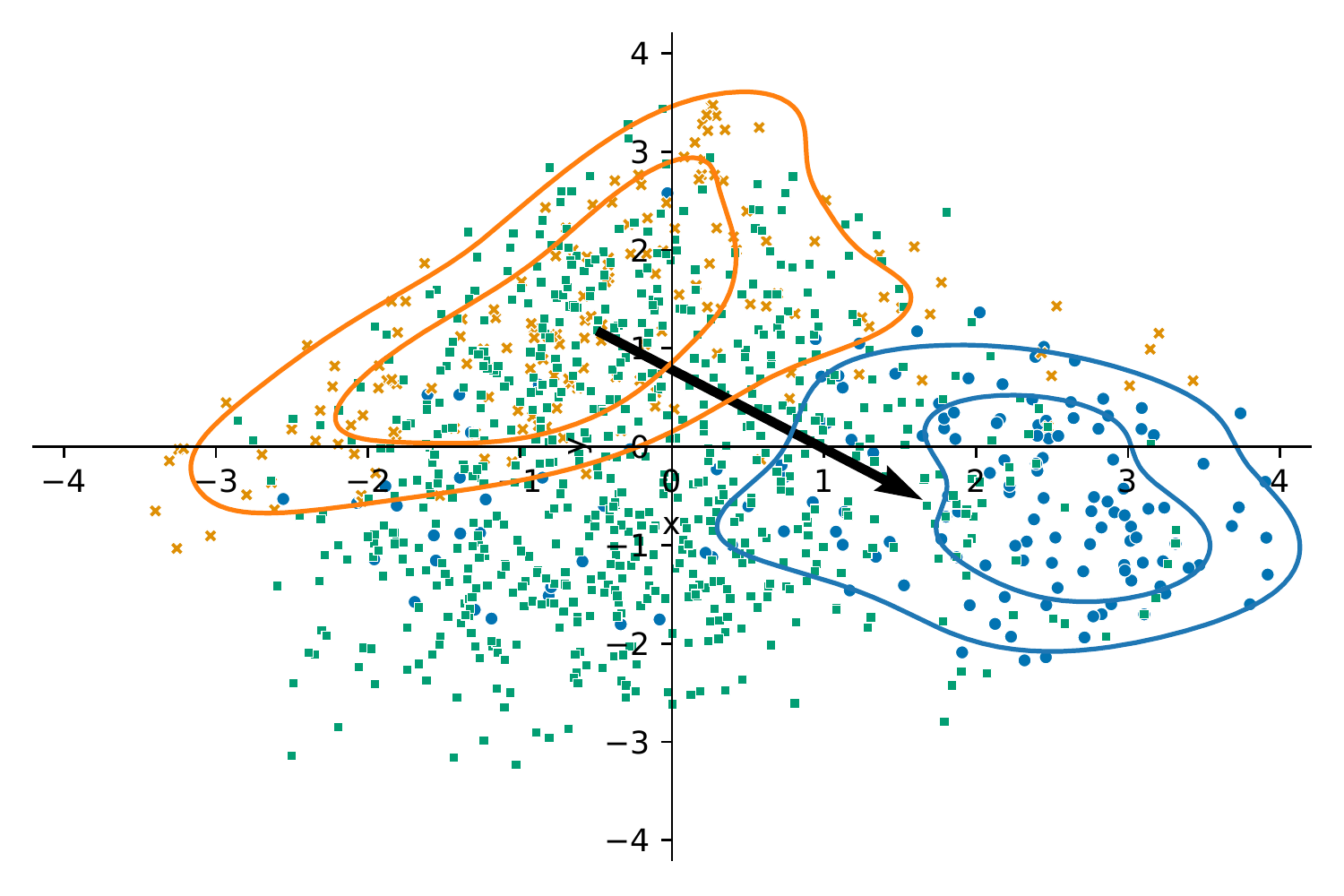}
        }
        \\
        \end{tabular}
    }
    
    % \el{was genau zeigt der pfeil?}\mrh{der pfeil geht von negativen  centroid zum positiven, dh beschreibt er die axis}\mrh{lass grad den plot neu erstellen, sodass es virtues und vices heisst} \el{ok, das ist gut! kannst du den pfeil auch weglassen? er verwirrt und wir haben nicht so viel platz um das gut u beschreiben}\mrh{glaub dann müssten wir validation of moral frame axes auch anpassen, ich würds so lassen}
    \caption{Axis of the five moral foundations. Each axis is created by the centroid of words assigned to virtues and the centroid of words for vices and surrounded by moral words associated with the other axes. The black arrow goes from the vices' centroid to the virtues' centroid and describes the axes. The high-dimensional space is reduced with Principal Component Analysis (PCA). All the axis point approximately in the same direction, which indicates that virtues are more similar to other virtues than to their corresponding vices, and vice versa. A kernel density estimation of the underlying point cloud is used for the colored contours.
    % Contours are created using kernel density estimation of the underlying point cloud.
    }
    \label{fig:frame_system}
\end{figure*}

% \section{Moral-based Frame Identification\mrh{improve section name: }}
\section{Characterization of Moral-based Framing}
\label{s:methods}
In the following, we describe our approach to investigate moral-based differences in the framing of tweets.

\subsection{Capturing Moral Frames} 
\label{ssec:foundations}

In our work, similar to~\citet{mokhberian2020moral}, we capture moral frames by combining the FrameAxis approach introduced in \citet{kwak2020frameaxis} with a  dictionary of moral values.
FrameAxis enables the quantification of framing of a particular text using \emph{semantic axes}~\cite{kwak2020frameaxis}. It is built upon the SemAxis approach~\cite{an2018semaxis}, which defines semantic axes by the difference of opposing word pairs using their word embeddings in the vector space.
FrameAxis learns in an unsupervised way by estimating the contribution of each word towards the target axis. The contribution per word is defined as the cosine similarity between its word embedding and the target axis in the vector space.
For all contributions of every word in a given document, we calculate the \emph{frame bias} and \emph{frame intensity} of a moral frame. The frame bias corresponds to the mean of the contributions and the frame intensity to the variance of the contributions in relation to the baseline frame bias of the corpus. The latter denotes the mean of frame biases over the whole corpus. 

As a dictionary of moral values, we use the Moral Foundation Dictionary version 2 (MFD-2)~\cite{frimer2017moral}. It is an extension of a moral values dictionary developed by~\citet{graham2009liberals} and consists of prototypical words to \emph{moral foundations}. Moral foundations are described in the moral foundation theory (MFT) as factors that guide emotional and ethical reactions to various social situations. MFT describes five foundations in the form of virtues and vices: (i) care/harm, i.e., the dislike for others' suffering, (ii) fairness/cheating, i.e., dislike of cheating, (iii) loyalty/betrayal, i.e., loyalty, (iv) authority/subversion, i.e., respect for authority, and (v) sanctity/degradation, i.e., concerns with purity~\cite{mokhberian2020moral}. 

The Moral Foundation Dictionary MFD-2 assigns words to virtues and vices. As virtues and vices are opposing moral values, we use them as poles to create \emph{moral frame axes}. Then, for each pole, we associate its words with word embeddings, i.e., the $300$-dimensional GloVe representation~\cite{pennington2014glove} trained on $840$ billion tokens and calculate their centroids for virtues and vices. Each pair of virtue and vice centroid forms a semantic axis, i.e., \emph{moral frame axis}, that we use for FrameAxis instead of individual words.
For each axis, we extract the frame biases and intensities per tweet by aggregating its words' contributions (i.e., the cosine similarity with the axis) towards the corresponding moral value. Please note that we name axes using the name of the morals' virtues in the remainder of this paper, e.g., the care axis.

\subsection{Validation of Moral Frame Axes}
\label{ssec:validate}
We define four properties of the word embedding space to investigate the validity of the moral frame axes. 
\emph{P1:~All axes should be close to the zero point.} Note that each axis is dividing a moral space into a positive and a negative part. \emph{P1} prohibits the dominance of one pole (i.e., the pole closer to the zero point) that could be caused by an association of an overwhelming majority of words. 
\emph{P2:~The words associated with a pole should be semantically closer to each other than to words of the opposite pole.}
If words are added to or removed from an axis, then \emph{P2} ensures its stability.
\emph{P3:~The orientation of axes should not oppose.}
Adherence to \emph{P3} allows the axes to be combinable and form a meta-axis for virtues and vices, e.g., care virtues are closer to fairness virtues than fairness vices. 
\emph{P4:~The orientation of axes should differ in the hyperspace.} We expect the axes to be orthogonal to a certain degree. A violation of \emph{P4}, i.e., two axes are pointing directly in the same direction, suggests that these axes likely relate to the same concept and could be combined.

A visual analysis of the moral frame axes (see Figure~\ref{fig:frame_system}) shows the first two principal components of word embeddings using probabilistic Principal Component Analysis (PCA), moral frame axes, and up to three density regions for virtues and vices using a kernel density estimation, which has a lowest level threshold of~$33\%$. 
Results indicate all the four properties hold, e.g., all the axes point in the same direction. Due to some ambiguous words, there is some overlap in the projected point clouds (e.g., unharmed). Furthermore, some words (e.g., wounds) belong to both poles, i.e., virtue and vice in the dictionary. In addition to the visual depiction, we also perform the validation numerically\footnote{We provide the code, plots and examples of this research at:\\ \url{https://github.com/socialcomplab/icwsm21-framing}\label{github}}.
% \footnote{Code and plots will be available in the non-blinded version.}. % on GitHub}.

\subsection{Validation on Annotated Tweets}
\label{ssec:evaluate}
To validate our approach, we perform classification of moral frames
similar to \citet{mokhberian2020moral} on the Twitter dataset provided by \citet{hoover2020moral}, which is annotated with virtues and vices. 
We conduct our experiments using a logistic regression classifier with the MFD-2 dictionary.
Table~\ref{tab:reproduce} contains the results of this experiment, and a comparison of our results with the results of~\citet{mokhberian2020moral}. 
% Overall, we observe similar results as Mokhberian et al., but also we find deviations in several  moral frames.
While we observe similar results as Mokhberian et al., we find that the use of MFD-2 improves the F1-score on care, fairness, and loyalty, but performs worse on authority and sanctity. In terms of accuracy, we achieve a higher performance on care and loyalty using MFD-2, but a lower performance on fairness, authority, and sanctity.
% From this experiment, w
We  conclude that the classifier accurately captures moral frames in tweets.

\begin{table}[ht]
    \centering
    \begin{tabular}{lrrrr}
    \toprule
    {} &  Precision &  Recall &  F1-score &  Accuracy \\
    \midrule
    \multicolumn{5}{c}{Reproduced with MFD-2} \\
    \hline
    Care &                   0.828 &                0.827 &                  \textbf{0.827} &     \textbf{0.827} \\
    Fairness &                   0.729 &                0.728 &                  \textbf{0.728} &     0.728 \\
    Authority &                   0.754 &                0.754 &                  0.754 &     0.754 \\
    Loyalty &                   0.895 &                0.889 &                  \textbf{0.891} &     \textbf{0.889} \\
    Sanctity &                   0.881 &                0.880 &                  0.880 &     0.880 \\
    % AVG  &                   0.817 &                0.816 &                 0.816 &     0.816 \\
    \midrule
    \multicolumn{5}{c}{Original with MFD-1} \\
    \hline
    Care      & 0.746 & 0.768 & 0.734 & 0.768 \\
    Fairness  & 0.662 & 0.774 & 0.681 & \textbf{0.774} \\
    Authority & 0.808 & 0.875 & \textbf{0.817} & \textbf{0.875} \\
    Loyalty   & 0.802 & 0.873 & 0.816 & 0.873 \\
    Sanctity  & 0.910 & 0.935 & \textbf{0.908} & \textbf{0.935} \\
    % AVG$^{*}$       & 0.771 & 0.822 & 0.775 & 0.822 \\
    \bottomrule
    \end{tabular}
    \caption{Results of classification of moral frames on the annotated Twitter corpus. The performance of MFD-2 is compared with the results from~\citet{mokhberian2020moral}.
    }
    \label{tab:reproduce}
\end{table}

\section{Experiments and Results}
\label{s:results}
We perform experiments on two datasets: firstly, in tweets in the English language created by US-based politicians, which we gathered based on the Twitter user list provided by \citet{barbera2015tweeting}, and secondly, in German-speaking tweets that contain COVID-19 related content created by followers of the leaders of the five major Austrian parties. Our selection of datasets is motivated by their differences in contextual attributes, concretely their language (i.e., English vs. German), topics (i.e., various topics vs. COVID-19-related topics), account type (i.e., politicians vs. followers of top politicians), and distribution of political parties (i.e., two-party system in the US vs. multi-party system in Austria).

\subsection{Datasets}
For the US Twitter dataset, 
we collect the most recent tweets of democrats and republicans using the party-associated Twitter handles~\cite{barbera2015tweeting}. The resulting dataset consists of $1,388,198$ tweets, i.e., $704,392$ tweets from $243$ democratic (D) and $683,806$ from $252$ republican (R) accounts. We label the tweets according to the account owner's party affiliation. 

For the Austrian Twitter dataset,
we manually extract the Twitter handles of the five major Austrian parties' lead politicians, i.e., @BMeinl for the liberal party (NEOS), @WKogler for the green party (Greens), @norbertghofer for the national-focused freedom party (FP\"O), @rendiwagner for the social-democratic party (SP\"O), and @sebastiankurz for the conservative people's party (\"OVP).
Then, we collect the most recent tweets of followers and labeled each tweet of the follower with the politician they follow. To avoid mutual labels, we restrict our collection to users that follow only one of the five accounts. Besides, we only consider tweets that contain COVID-19 related hashtags (e.g., \#Corona). This results in a collection of $22,205$ tweets, i.e., $17,230$ tweets labeled with @sebastiankurz, $2,090$ labeled with @WKogler, $1,164$ labeled with @rendiwagner, $901$ labeled with @BMeinl, and $820$ labeled with @norbertghofer.

We normalize the tweets in both datasets and (i.e., lowercase, removing URLs, punctuation), remove stopwords, and apply tokenization before extracting the frame biases and intensities for training a logistic regression classifier\textsuperscript{\ref{github}}.

\begin{table*}[!th]
% \qquad
\subfloat[Moral frames in US politics. Democrats (D) and republicans (R) differ most in terms of frame intensities (in bold).]
{
\centering
\begin{tabular}{ll|rr}
\toprule
  \multicolumn{2}{l|}{\textbf{Moral Frames}}  & D         & R         \\
    %   & Morals     & Democrats         & Republicans         \\
\midrule
\parbox[t]{3mm}{\multirow{5}{*}{\rotatebox[origin=c]{90}{Bias}}}   
  & Care       &  2.505  &  3.791  \\
  & Fairness   &  2.130  &  1.115  \\
  & Authority  & -0.385  &  2.343  \\
  & Loyalty    & -1.419  & -5.269 \\
  & Sanctity   &  0.476  &  2.102  \\
\cline{1-4}
\parbox[t]{3mm}{\multirow{5}{*}{\rotatebox[origin=c]{90}{Intensity}}} 
  & Care       &  \textbf{9.701}  & -0.634 \\
  & Fairness   &  \textbf{4.376}  & -8.154 \\
  & Authority  & -3.453  & -6.329 \\
  & Loyalty    &  0.166  &  \textbf{9.956}  \\
  & Sanctity   & -2.967  &  \textbf{3.261} \\
\bottomrule
\end{tabular}
    \label{tab:unites_states_coef}
}
\quad
\subfloat[Moral frames in Austrian politics. Frame biases are distinct between the followers of the party leaders, whereas intensities are very small in comparison. 
Minimum and maximum of frame biases per moral are in bold. Frame bias in fairness exhibits the greatest difference.
% Extreme frame biases (i.e., minimum and maximum) per moral value are in bold
% with fairness exhibiting the greatest difference.
]{
\centering
\begin{tabular}{rrrrr}
\toprule
@BMeinl    & @WKogler   & @norbertghofer & @rendiwagner & @sebastiankurz \\
% BMeinl    & WKogler   & norbertghofer & rendiwagner & sebastiankurz \\
% NEOS & Greens & FP\"O & SP\"O & \"OVP \\
\midrule
   -0.788 & -0.463 & \textbf{-4.682}     &  2.141    &  \textbf{6.931}      \\
   -1.375 & \textbf{12.494} & -9.165     &  2.408    & \textbf{-13.408}    \\
   -0.035 & \textbf{-1.078} & -0.366     &  \textbf{2.561}    & -0.291     \\
   -0.078 & -0.998 &  \textbf{2.627}     & \textbf{-7.987}    &  0.649      \\
   \textbf{-3.673} & -0.457 & \textbf{15.645}     & -0.145    &  0.458      \\
\cline{1-5}
   0.077  & -0.039  &  0.001      & -0.010    & -0.011     \\
   0.072  & -0.046  &  0.038      & -0.066    &  0.018      \\
   0.008  &  0.002  & -0.009      &  0.007    & -0.003     \\
   0.003  &  0.003  & -0.003      &  0.001    & -0.009     \\
   0.104  & -0.008  & -0.023      & -0.020    & -0.034    \\
\bottomrule
\end{tabular}
    \label{tab:austrian_coefs}
}
\caption{Reported results correspond to the coefficients of the logistic regression classifier.}
\label{tab:coefs}
\end{table*}

\subsection{Moral-based Framing in US-based Tweets}
\label{ssec:american}
We group the tweets by parties and report the coefficients of the logistic regression classifier in Table~\ref{tab:unites_states_coef}.
The frame biases do not deviate considerably and, in general, share the same direction on all moral frames but on authority, which is positive for republicans and negative for democrats. We observe that democrats score higher in fairness and lower in sanctity, whereas republicans score higher in the frame bias for care and exhibit a high negative score in the frame bias for loyalty. Concerning the frame intensities, we observe opposing and more distinct  results. The frame intensity for care is much higher for democrats, and conversely, the frame intensity for loyalty is higher for republicans. The frame intensities on fairness and sanctity agree with their corresponding frame biases, i.e., fairness has a higher frame intensity for democrats, while sanctity has a higher frame intensity for republicans. We find that our observations are congruent with \citet{graham2009liberals}, i.e., liberals are predominantly associated with care and fairness.

\subsection{Moral-based Framing in Austrian-based Tweets}
\label{ssec:austrian}
To investigate differences in moral-based framing in the Austrian Twitter dataset, we first translate the content of the MFD-2 dictionary. To that end, we use a list of sample translations of positive and negative valence words~\cite{weichselbaum2018implicit} and two sets of word embeddings, i.e., one for English and one for German~\cite{grave2018learning}. Using a translation matrix estimated from the valance word translations, we translate the words of the MFD-2 to similar words in German in terms of their word embeddings. We see that the top words seem to be congruent with the moral values, e.g., top translation of authority being \emph{Befehl} -- \emph{command}, but also observe words of opposite moral values in their vicinity, e.g., harm having \emph{Schadenfreude} -- \emph{malicious joy} as the second, and \emph{Freude} -- \emph{joy} and third nearest neighbor. Such inconsistencies are expected since we previously established that some words are neither clearly associated with virtues nor vices.

Then, we group the tweets by followers of Austrian party leaders and report the coefficients of the logistic regression classifier in Table~\ref{tab:austrian_coefs}. We find substantial differences in frame biases between the tweets of the groups, but not in their frame intensities.
The reported frame biases reaffirm the parties' public perception, with fairness having a stronger association with left parties (with @WKogler followers being the highest), while sanctity is predominantly associating with right parties (i.e., the highest for @norbertghofer followers).
Noteworthy, the followers of @sebastiankurz have the lowest association with fairness, which might indicate a contention point between the viewpoints of the governing coalition, i.e., the \"OVP (@sebastiankurz) and Greens (@WKogler). Moreover, the results show that @sebastiankurz followers are mostly associated with care, a moral frame that is prevalent in the government's COVID-19 information campaign through the slogan \emph{"Schau auf dich - schau auf mich"}, which translates to \emph{"take care of you - take care of me"}.
Followers of @rendiwagner, who is also a  scientist and epidemiologist, are associated with authority. We suspect that is the result of her emphasizing to listen to doctors and experts. For followers of @BMeinl, all frame biases are negative, which we relate to the party being an opposition party arguing against government COVID-19 policies.
In summary, we find differences in the moral framing of the tweets on COVID-19 of the followers of the party leaders that reflect the ideology and messages of the corresponding political parties.

\section{Conclusion}
In conclusion, our experimental results show that the moral framing in the tweets of US-based politicians and the tweets of the followers of Austrian politicians is congruent with the public perception of the political parties. In the tweets from US-based politicians, we find that democrats are associated with high frame intensity in care and fairness, whereas high frame intensity in loyalty and sanctity is associated with republicans. In the tweets from followers of the five major Austrian parties' leaders, we find that high frame bias in fairness is mostly associated with followers of the green party's leader, while high frame bias in sanctity predominantly indicates followers of the freedom party's leader. Besides, we find that followers of the ruling conservative party's leader have a notable frame bias towards care in the case of COVID-19-related tweets. We attribute this to the followers' adoption of the framing of the conservative COVID-19 slogan that stresses caring. 
From a methodological perspective, our experiments show that the use of the extended moral foundations dictionary MFD-2 increases the accuracy of moral frame characterization. % and that it can be generalized to German using sample translations of positive and negative valence words in combination with word embeddings. 
%to studying moral-based differences in the framing of political tweets 
% From a methodological perspective, our experiments show that the use of the extended moral foundations dictionary MFD-2 increases the accuracy of moral frame characterization.

% \paragraph{Limitations}
We recognize several limitations of our work: our analysis is restricted to two specific political Twitter datasets. We chose these datasets, as the interpretation of results requires the researchers' domain understanding and language skills. Through making a validity analysis of the approach, we aimed to mitigate the potential impact of constraints. 
Also, since we did not filter out retweets, $63$ tweets in the Austrian dataset are from the political party leaders. % themselves.
% Also, in the Austrian dataset, $63$ tweets of the Austrian party leaders are in the dataset since we did not filter out retweets. 
% However, only $63$ direct retweets of the party leaders are in the dataset. 
% Also, in the Austrian dataset, also, tweets of the politicians can be in the dataset since we did not filter out retweets. However, only $63$ direct retweets of the party leaders are in the dataset. 
% \mrh{haben da anscheinend vorbei geredet, 39 waren nur für kurz. insgesamt sind es 63, ist aber trozdem sehr gering}
% \el{However, we find that this is only the case for @sebastiankurz, for whome 39 direct retweets are in the dataset.}  %Nevertheless, it is not known whether the approach generalizes to other contexts e.g., other languages or topics.

% \paragraph{Future Work}
For future work, we aim to research  the interplay of frame bias and intensity in more detail. We will also study how followers engage with moral frames shared by politicians and if they are more prevalent in retweets or comments.

%These studies will not only be limited to Twitter but also consider traditional news articles. 
% We will also examine the relationship between framing and polarization. % in more detail.

\bibliography{icwsm21}

\end{document}